# Contact angle measurement in lattice Boltzmann method

Binghai Wen[1,2], Bingfang Huang[1], Zhangrong Qin[1], Chunlei Wang[2,*], Chaoying Zhang[1,†]

[1]Guangxi Key Lab of Multi-source Information Mining & Security, Guangxi Normal University, Guilin 541004, China

[2]Division of Interfacial Water and Key Laboratory of Interfacial Physics and Technology, Shanghai Institute of Applied Physics, Chinese Academy of Sciences, Shanghai 201800, China

Corresponding Authors：

*E-mail: wangchunlei@sinap.ac.cn;    †E-mail: zhangcy@gxnu.edu.cn;

Contact angle is an essential characteristic in wetting, capillarity and moving contact line; however, although contact angle phenomena are effectively simulated, an accurate and real-time measurement for contact angle has not been well studied in computational fluid dynamics, especially in dynamic environments. Here, we design a geometry-based mesoscopic scheme for on-the-spot measurement of the contact angle in the lattice Boltzmann method. The measuring results without gravity effect are in good agreement with the benchmarks from the spherical cap method. The performances of the scheme are further verified in gravitational environments by simulating sessile and pendent droplets on smooth solid surfaces and dynamic contact angle hysteresis on chemically heterogeneous surfaces. This scheme is simple and computationally efficient. It requires only the local data and is independent of multiphase models.

## 1. Introduction

Contact angle, which indicates the wettability of a solid surface by a liquid, is a characteristic quantity in a great amount of wonderfully natural phenomena and significantly industrial applications, such as capillarity, microfluidics, nanotechnology, moving contact line, coating technology, etc[1, 2]. Essentially, both of the static and dynamic contact angles should be measured at contact line on the microscale[1]. Experimenters have developed all kinds of methods to determine contact angle. Bigelow *et al*. set up the most widely used technique, which utilized a telescope-goniometer to directly measure the tangent angle at the three-phase contact point on a sessile droplet profile[3]. Angles measured in such a way are often quite close to advancing contact angles. Equilibrium contact angles can be obtained through the application of well-defined vibrations[4]. Extrand and Kumagai studied the contact angle hysteresis on a variety of polymer surfaces by using an inclined plate method, in which a sessile droplet locates on a inclined plate and both of the advancing and receding contact angles are simultaneously obtained[5]. Kwok *et al*. used a motor-driven syringe to control the rate of liquid addition and removal to study advancing, receding, or dynamic contact angles[6]. Besides observing a sessile droplet on a solid sample, a telescope-goniometer is also necessary in other contact angle measurements. The captive bubble method forms an air bubble beneath the solid sample, which is immersed in the testing liquid[7]. The contact angle formed by the air bubble in liquid can also be directly measured. The tilting plate method applies a solid plate with one end immersed in the liquid and forms a meniscus on both sides of the plate[8]. The plate inclines slowly until the meniscus becomes horizontal on one side of the plate and then the angle between the plate and the horizontal is the contact angle. It is generally recognized that the direct measurement of drop contact angles with a telescope-goniometer can yield an accuracy of approximately ±2°[9]. The Wilhelmy balance method is another type of popular scheme to measure contact angle[10]. A solid sample is manipulated to immerse into or emerge from the wetting liquid. The task of measuring an angle is reduced to the measurements of the weight and length, which can be performed with high accuracy and without subjectivity. The method is also suitable to measure dynamic contact angle and hysteresis, because the three-phase line can be in wholesale motion assuring achievement of maximal advancing and minimum receding contact angles[11]. The experimental measurements of the contact angle promote to investigate the surface tensions and wetting mechanisms of the solid surfaces, especially interpretation of contact angles in terms of surface energetics of solids[9].

Numerical simulation has been developed into an effective way to research fluid flow and is expected to provide more rich details than experiments. In computational fluid dynamics, there are usually several methods available to measure a contact angle. The measurement of the drop image by a goniometer is relatively rough and subjective after the images are generated from the simulation data[12]. In a more accurate way, a graphical analysis can be applied to obtain the contact angle from the image[13]. Since the fluid images have to be exported before the measurement, these methods are time-consuming and cannot serve as an on-the-spot measurement. Ignoring the gravity effect, a droplet holds a perfect spherical cap on a horizontal solid surface owing to the surface tension. The contact angle can be accurately calculated based on the measurement of the height and bottom width of the droplet[14, 15]. This scheme is referenced here as the spherical-cap method. On chemically striped patterned surfaces, the contact line is corrugativus, the contact angle can be determined using the height of the droplet and the radius of curvature, which fits the droplet profile[16]. It is more complex in molecular dynamics simulations. Since the drop size reduces to nanoscale, there is not a steady interface between gas and liquid. The drop contours have to be fitted by a least square technique[17, 18]. Although these theoretical methods are simple and easy to implement, they are limited in a zero-gravity equilibrium environment. For diffuse-interface simulations, Ding *et al*. proposed a geometric formulation of wetting condition based on the gradient of the volume fraction for binary fluid flows, by which the prescribed contact angle can be correctly obtained[19]. Lee *et al*. improved the accuracy of the contact angle boundary condition as well as its numerical stability by a characteristic interpolation[20]. Considering to the relaxation of dynamic contact angle, Dong further extended the contact-angle boundary conditions to simulate dynamic wall-bounded gas/liquid flows with large density ratio[21]. Leclaire *et al*. imposed the desired contact angle at the boundary as a Dirichlet boundary condition and then studied immiscible two-phase pore-scale imbibition and drainage in porous media [22, 23]. As for these contact angle conditions, the main efforts were focusing on the wetting boundary constraints, but not on the evaluations of contact angles, even more not on the calculation of the dynamic contact angle. Moreover, these imposing procedures of contact-angle boundary conditions are computationally complex and nonlocal. Especially, they involve the intervention to the evolution of flow field. Therefore, a simple, exact and on-the-spot measurement of contact angle is meaningful for the numerical investigation of wetting phenomena.

Essentially, contact angle is a geometrical concept. Only for some special cases, such as a sessile droplet at zero-gravity mechanical equilibrium on a horizontal surface, the contact angle can be theoretically explained by Young's equation. In a dynamic or nonequilibrium environment, the contact angle should be measured through a geometrical method. The lattice Boltzmann method has developed into an alternative tool to model multiphase flow systems, and been successfully applied to many of the fields related to the surface wetting science and engineering application [15, 24]. Its regular and mesoscopic lattices lay a foundation for efficient contact angle measurement. In this paper, we design a geometry-based mesoscopic scheme to measure the real-time contact angle. The various test cases with and without gravity are conducted to verify the proposed scheme. The computational results show that the scheme is simple, accurate and efficient.

The paper is organized as follows. In section 2, we introduce the chemical-potential multiphase lattice Boltzmann model. Section 3 proposes the mesoscopic contact angle measurement, whose computational accuracy is verified by comparing with the benchmarks. In section 4, a series of droplet deformations under gravity are simulated and the contact angles are measured by the proposed method. Section 5 is about the investigations of the contact angle hysteresis. Finally, section 6 concludes the paper.

## 2. Chemical-potential-based multiphase model

Numerical simulation of multiphase flow is one of the most successful applications for the lattice Boltzmann method (LBM)[15, 24]. Originating from the cellular automaton concept and kinetic theory, the intrinsic mesoscopic properties make LBM outstanding to model complex fluid systems involving interfacial dynamics[25-28] and phase transitions[29-37]. Discretized fully in space, time and velocity, the lattice Boltzmann equation with a single relaxation time can be concisely written as [38]

$$f_i(\boldsymbol{x}+\boldsymbol{e}_i,t+1) - f_i(\boldsymbol{x},t) = -\frac{1}{\tau}[f_i(\boldsymbol{x},t) - f_i^{(eq)}(\boldsymbol{x},t)], \qquad (1)$$

where $f_i(\boldsymbol{x},t)$ is the particle distribution function at lattice site $\boldsymbol{x}$ and time $t$, $\boldsymbol{e}_i$ is the discrete speeds with $i=0,...,N$, $\tau$ is the relaxation time, and $f_i^{(eq)}$ is the equilibrium distribution function

$$f_i^{(eq)}(\boldsymbol{x},t) = \rho\omega_i[1 + 3(\boldsymbol{e}_i\cdot\boldsymbol{u}) + \frac{9}{2}(\boldsymbol{e}_i\cdot\boldsymbol{u})^2 - \frac{3}{2}u^2], \qquad (2)$$

where $\omega_i$ is the weighting coefficient and $\boldsymbol{u}$ is the fluid velocity. The evolution of the LBE can be decomposed into two elementary steps, collision and advection:

$$\text{collision:} \quad \tilde{f}_i(\boldsymbol{x},t) = f_i(\boldsymbol{x},t) - \frac{1}{\tau}[f_i(\boldsymbol{x},t) - f_i^{(eq)}(\boldsymbol{x},t)], \qquad (3)$$

$$\text{advection:} \quad f_i(\boldsymbol{x}+\boldsymbol{e}_i,t+1) = \tilde{f}_i(\boldsymbol{x},t), \qquad (4)$$

where $f_i$ and $\tilde{f}_i$ denote pre-collision and post-collision states of the particle distribution functions, respectively. The mass density and the momentum density are defined by $\rho = \sum f_i$ and $\rho\boldsymbol{u} = \sum \boldsymbol{e}_i f_i$, respectively.

Considering a nonideal fluid system, the free-energy functional within a gradient-squared approximation is [30, 34, 39]

$$\Psi = \int(\psi(\rho) + \frac{\kappa}{2}|\nabla\rho|^2)d\boldsymbol{x}, \qquad (5)$$

where the first term of the integrand is the bulk free-energy density at a given temperature with the density $\rho$ and the second term gives the free-energy contribution from density gradients in a inhomogeneous system. The free energy

function in turn determines the diagonal term of the pressure tensor

$$p(\boldsymbol{x}) = p_0 - \kappa\rho\nabla^2\rho - \frac{\kappa}{2}|\nabla\rho|^2, \quad (6)$$

with the general expression of equation of state (EOS)

$$p_0 = \rho\psi'(\rho) - \psi(\rho). \quad (7)$$

The full pressure tensor can be written as

$$P_{\alpha\beta}(\boldsymbol{x}) = p(\boldsymbol{x})\delta_{\alpha\beta} + \kappa\frac{\partial\rho}{\partial x_\alpha}\frac{\partial\rho}{\partial x_\beta}. \quad (8)$$

where $\delta_{\alpha\beta}$ is the Kronecker delta.

For a van der Waals fluid, the chemical potential can be defined from the free energy density functional[39-41]

$$\mu = \frac{\partial\Phi}{\partial\rho} - \nabla\cdot\frac{\partial\Phi}{\partial(\nabla\rho)} \quad (9)$$

where

$$\Phi(\rho) = \psi(\rho) + \frac{\kappa}{2}(\nabla\rho)^2. \quad (10)$$

Then the chemical potential can be computed by the density and free-energy density

$$\mu = \psi'(\rho) - \kappa\nabla^2\rho \quad (11)$$

Substituting Eq. (7) and (11) into Eq. (8), a graceful relationship is obtained[42]

$$\nabla\cdot\vec{\boldsymbol{P}} = \rho\nabla\mu_\rho. \quad (12)$$

Thus, the nonideal force can be easily evaluated by the chemical potential avoiding the pressure tensor

$$\boldsymbol{F}(\boldsymbol{x}) = -\nabla\cdot\vec{\boldsymbol{P}}(\boldsymbol{x}) + \nabla\cdot\vec{\boldsymbol{P}}_0(\boldsymbol{x}) = -\rho\nabla\mu_\rho + \nabla\cdot\vec{\boldsymbol{P}}_0(\boldsymbol{x}). \quad (13)$$

where $\vec{\boldsymbol{P}}_0 = c_s^2\rho\vec{\boldsymbol{I}}$ is the ideal-gas equation of state. Then, the nonideal force acts on the collision process by simply increasing the particle momentum in $f_i^{(eq)}$ in terms of the momentum theorem. The velocity in Eq. (2) is replaced by an equilibrium velocity[43]

$$\boldsymbol{u}^{eq} = \boldsymbol{u} + \frac{\tau\boldsymbol{F}}{\rho}. \quad (14)$$

Correspondingly, the macroscopic fluid velocity is redefined by the averaged momentum before and after the collision

$$\mathbf{v} = \mathbf{u} + \frac{\mathbf{F}}{2\rho}. \tag{15}$$

This multiphase model satisfies thermodynamics and Galilean invariance [42]. It can work together with various equations of state to simulate all kinks of multiphase flows.

In this paper, we model the popular water-vapor system by the chemical-potential-based multiphase lattice Boltzmann method together with the famous Peng-Robinson (PR) EOS,

$$p_0 = \frac{\rho RT}{1-b\rho} - \frac{a\alpha\rho^2}{1+2b\rho-b^2\rho^2} \tag{16}$$

where the temperature function is $\alpha(T) = [1+(0.37464+1.54226\omega-0.26992\omega^2)\times(1-\sqrt{T/T_c})]^2$ with the acentric factor $\omega = 0.344$ for water. The parameters take $a = \frac{2}{49}$, $b = \frac{2}{21}$ and $R=1$, thus the critical temperature is $T_c = 0.07292$ and the critical density is $\rho_c = 2.65730$. The chemical potential of PR EOS is [42]

$$\mu = RT \ln\frac{\rho}{1-b\rho} - \frac{a\alpha}{2\sqrt{2}b}\ln\frac{\sqrt{2}-1+b\rho}{\sqrt{2}+1-b\rho} + \frac{RT}{1-b\rho} - \frac{a\alpha\rho}{1+2b\rho-b^2\rho^2} - \kappa\nabla^2\rho. \tag{17}$$

In an effort to relate the numerical results to the real physical properties, the reduced variables $T_r = T/T_c$ and $\rho_r = \rho/\rho_c$ are used in the following simulations.

## 3. Mesoscopic contact angle measurement
### 3.1. Chemical-potential boundary condition

Endowing a solid surface with a chemical potential, the wettability, namely the interaction between the fluid and the solid, can be well regulated. As shown in Fig. 1, a straight interface between fluid and solid locates on a row of lattice nodes (y=1), which are treated as fluid nodes. The distribution functions on these interfacial fluid nodes still collide and stream, and the bounce-back boundary condition is applied to make up those distribution functions from solid. During the evaluation of nonideal force, the densities of the solid nodes (y=0) must be estimated in order to calculate the density gradients by Eq. (15) on these interfacial nodes. A simple weighted average scheme of the neighbor fluid nodes is used here

$$\rho(x,0) = \frac{2}{3}\rho(x,1) + \frac{1}{6}\rho(x-1,1) + \frac{1}{6}\rho(x+1,1). \qquad (18)$$

A uniform chemical potential is assigned to the solid nodes (y=0) in order to specify the wettability of the solid surface. It will be used to evaluate the gradient of chemical potential in Eq. (20). By changing the chemical potential, the contact angle can be easily adjusted.

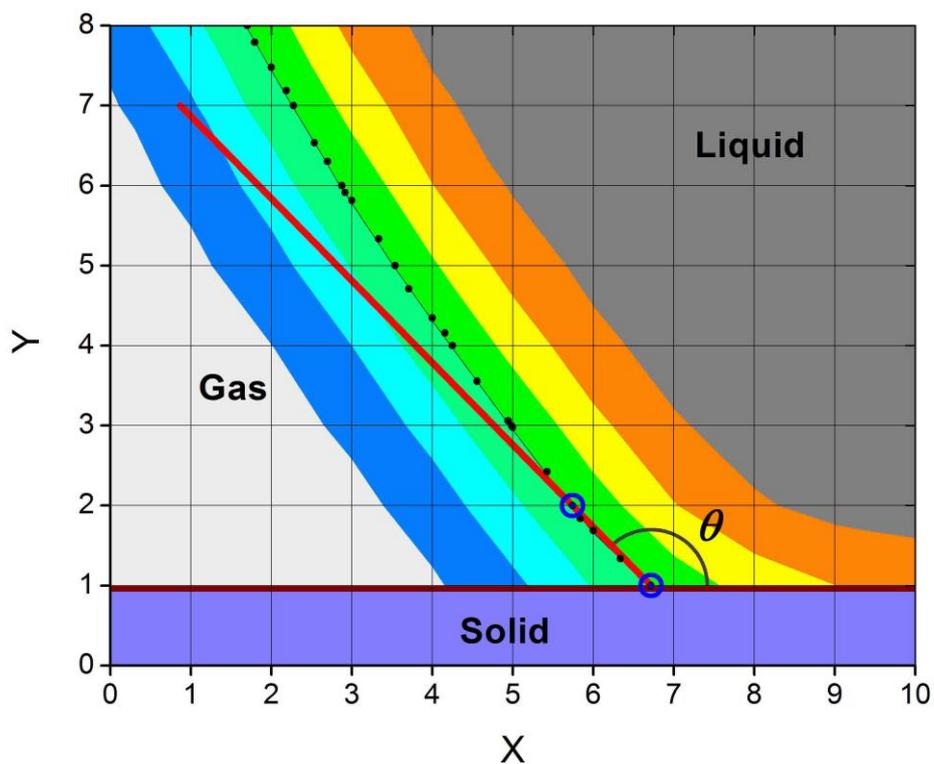

*Fig. 1. (color online) A schematic diagram to illustrate the chemical-potential boundary condition and the measurement of contact angle on mesoscopic scale. The small black spots are the intersections of droplet surface and gas-liquid links. The first point (y=1) and the second point (y=2) are marked by blue circles. The red line represents the tangential line of the droplet at the triple-phase contact point.*

## 3.2. Contact angle measurement

Aiming to calculate contact angle with high resolution on mesoscopic scale, two questions must be solved at first. One is where the gas-liquid interface locates exactly and the other is how to get the stable extreme value of contact angle at the triple-phase contact point. Fortunately, once the questions are raised, the answers emerge naturally. The drop surface must intersect with the gas-liquid links, each of which connects a gas node, whose density is smaller than the mean density, and a liquid node, whose density is greater than the mean density. A linear interpolation is the best candidate to locate the accurate position of the drop surface on a gas-liquid link,

$$\boldsymbol{x} = \boldsymbol{x}_g + \frac{\rho_m - \rho(\boldsymbol{x}_g)}{\rho(\boldsymbol{x}_l) - \rho(\boldsymbol{x}_g)} \boldsymbol{e}, \qquad (20)$$

where $\boldsymbol{x}_g$ and $\boldsymbol{x}_l$ are the gas and liquid lattice nodes connected by a gas-liquid link, $\rho_m$ denotes the average of the gas and liquid density, and $\boldsymbol{e}$ represents the link direction from gas to liquid. It is worth to notice that $\rho(\boldsymbol{x}_g)$ and $\rho(\boldsymbol{x}_l)$ in the transition region may be not equal to the coexistence densities $\rho_g$ and $\rho_l$ respectively. Consequently, a series of discrete surface points of the droplet are obtained as shown in Fig. 1 and the point on the solid surface (y=1) is set as the triple-phase contact point on mesoscopic scale. Now, we need to select another point to determine the contact angle. Among the nearby points, we find that the point (y=2) gives the most accurate and stable angle value for most of contact angle range. The reasons are easy to understand: for one thing the point (y=2) keeps an appropriate distance to the triple-phase contact point (y=1); for another, both of the two points are on the horizontal gas-liquid link and then receive similar calculation errors induced by Eq. (20). As for emphasis in Fig. 1, both of the first point (y=1) and the second point (y=2) are marked by blue circles. Thus, the contact angle $\theta$ can be defined by the horizontal positive direction and the ray from the first point to the second point, which is red in Fig. 1. When the contact angle is larger than $170^o$, the nearer second point has to be used to reduce the angle error. Especially, when the first point is no longer on the solid surface and the droplet keeps touching to the solid, the contact angle reaches $180^\circ$, which is a delicate state in a zero-gravity environment. It should be noted that the mesoscopic scheme of contact angle measurement is independent of the chemical-potential-based multiphase model.

### 3.3. Mesoscopic morphology

We select the popular water-vapor system to demonstrate the present model by using the famous Peng-Robinson (PR) EOS. Ignoring the gravity effect, a sessile droplet on a solid surface with various specified chemical potentials is simulated to achieve different surface wettabilities. The computational domain is a rectangular with the length $Dx = 300$ and the width $Dy = 100$. The density field is initialized as follows [44]:

$$\rho(x, y) = \frac{\rho_g + \rho_l}{2} + \frac{\rho_g - \rho_l}{2} \tanh\left[\frac{2(r - r_0)}{W}\right], \tag{19}$$

where $\rho_g$ and $\rho_l$ are the two-phase coexistence densities obtained by the Maxwell equal-area construction, $W = 5$ is the initial interface width, $r_0$ is the drop initial radius and $r = \sqrt{(x - x_0)^2 + (y - y_0)^2}$, in which $x_0 = Dx/2$ and $y_0 = r_0 + 1$. The temperature takes $T_r = 0.8$ and the drop radius is $r_0 = 30$. Each case performs 100,000 time steps of evolution to achieve its equilibrium state. The periodical boundary condition is applied to the left and right sides of the flow field, while the chemical potential boundary condition is used for the top and bottom sides. For the top side, the value of chemical potential is optional because the droplet never touches it. Typically, the chemical potential on a top side node takes the same value as that of the neighbor fluid nodes, which is always gas phase in the simulations. For the bottom side, a set of chemical potentials are specified to the nodes.

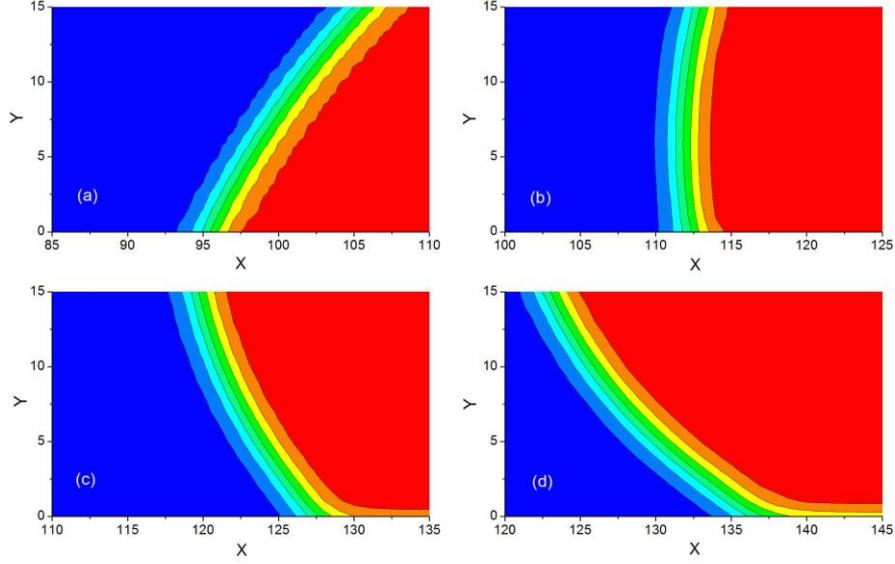

*Fig. 2. (Color online) The mesoscopic morphology of contact angles: (a) $\theta = 64^o$ at CP = 0.2, (b) $\theta = 100^o$ at CP = 0.3, (c) $\theta = 135^o$ at CP = 0.4, (d) $\theta = 155^o$ at CP = 0.45.*

The equimolar dividing surface is usually considered as the theoretical interface between gas and liquid[39]; however, its determination has to perform a heavy computation. Here, we define the drop surface as a contour line where the density is equal to the mean density of the gas phase and liquid phase. Fig. 2 further draws the mesoscopic morphology of different contact angles. It is clearly shown that the droplet surfaces directly fall on the solid surface. This feature offers an opportunity to accurately measure the contact angle at mesoscopic scale.

## 4. Droplets without gravity

Without gravity, a droplet on a horizontal solid surface will form a perfect spherical cap. If the base and height of the droplet are $L$ and $H$, the radius of the droplet is calculated by $R = (4H^2 + L^2)/8H$ and then the spherical cap method evaluates the contact angle by the formula $\tan\theta = L/2(R-H)$. Since the measurements of the droplet base and height are convenient and accurate, the spherical cap method can be used as a benchmark to verify the proposed mesoscopic measurement of contact angle. The computational domain is extended to $Dx = 500$ and $Dy = 200$, and the drop radius takes $r_0 = 40$. The droplets on a horizontal solid surface at two temperatures $T_r = 0.7, 0.8$ are simulated by using PR EOS and the numerical contact angles are compared with those from the spherical cap method. As shown in Figs. 3 and 4, with the growth of the chemical potential of the solid surface, the simulating contact angles smoothly increase and are in good agreement with the benchmarks. Therefore, the chemical-potential multiphase model is competent to simulate the contact angle phenomena and the mesoscopic measurement is accurate. Remarkably, the contact angle increases almost linearly along with the chemical potential, hence it is very convenient to regulate the contact angle for practical requirements in the present chemical-potential multiphase model.

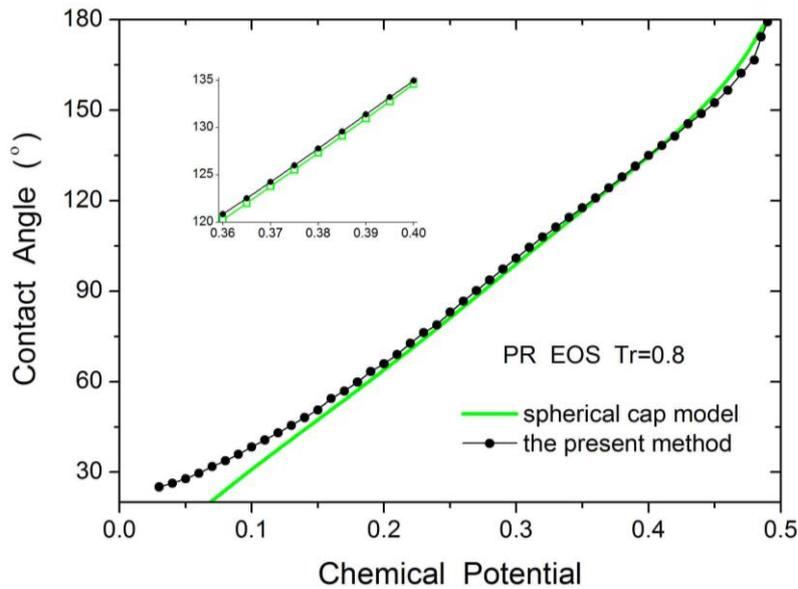

*Fig. 3. (Color online) Contact angle increasing with the chemical potential of the solid surface at the reduced temperature 0.8.*

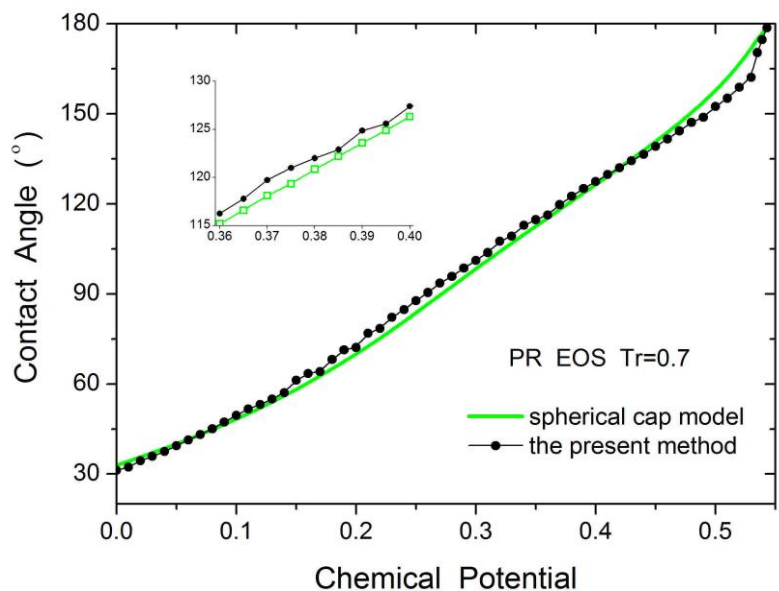

*Fig. 4. (Color online) Contact angle increasing with the chemical potential of the solid surface at the reduced temperature 0.7.*

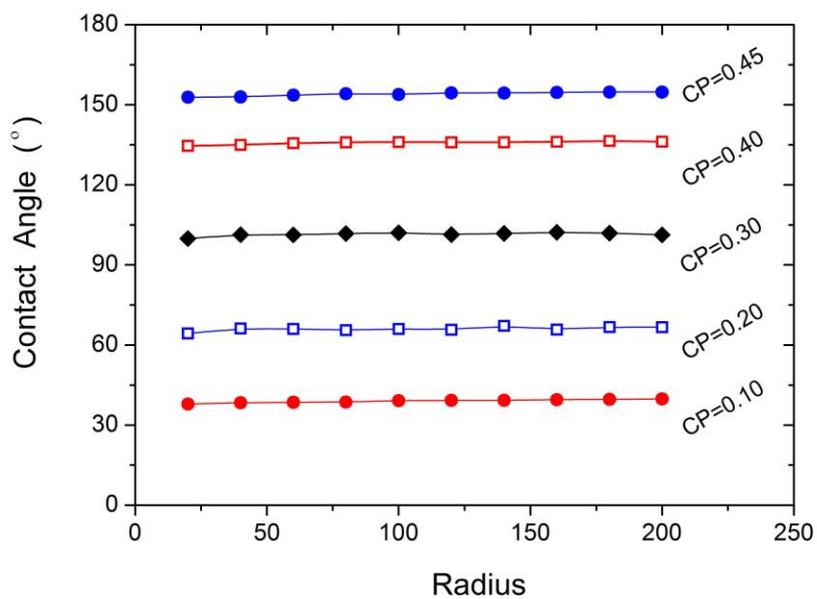

*Fig. 5. (Color online) Contact angles of the droplets with a series of initial radiuses on different solid surface.*

Theoretically, without gravity, the contact angles will remain the same for different sizes of drops on the same solid surface at a given temperature. This can be used to examine the effect of drop size on the contact angle measurement. We simulate a series of droplets, whose initial radiuses gradually increase from $r_0 = 20$ to $r_0 = 200$. It can be seen clearly in Fig. 5 that for all kinds of solid surfaces, the contact angles of the droplets with different radiuses are highly consistent. Therefore, the present mesoscopic measurement of contact angle is independent of the size and resolution of the drop.

**5. Droplet deformations under gravity**

Theory and experiments support that gravity has no effect on the equilibrium contact angle of a droplet on a smooth homogeneous surface[45]. Naturally, with the effect of gravity, a droplet on a solid surface will deform deviating from an ideal spherical cap: a sessile droplet is squashed and forms an ellipsoidal cap, while a pendent droplet is stretched and forms a protuberant cap. Thus, the spherical cap method will produce serious deviations, which grow fast along with the drop size. The sessile droplets and pendent droplets are separately simulated. The computational domain keeps $Dx = 500$ and $Dy = 200$, the drop radius takes $r_0 = 50$ and the temperature is $T_r = 0.8$. The drop density is 1 g/cm$^3$, the gravity acceleration is $|G| = 980$ cm/s$^2$. The droplet on lattice unit is mapped to the macroscopic droplet by dimensional transformation. With the growth of the macroscopic drop size, the effect of gravity is increasingly pronounced. The initialization of flow field is the same as that in section 3.2. After 10,000 time steps of free evolution, the gravity force is gradually exerted on the fluid, both gas and liquid, and then the system evolves till 100,000 time step.

## 5.1. Sessile droplets

Owing to the gravity effect, a sessile droplet will be flattened, its base extends and its height lowers. The contact angle calculated by the spherical cap method will reduce unsurprisingly. We simulate a series of droplets, whose diameters change from 0.01 cm to 0.5 cm. These droplets are located on three solid surfaces with the chemical potentials 0.2, 0.3 and 0.4, respectively. The macroscopic diameter 0 equals to the situation of zero gravity. Fig. 6 illustrates that when the initial drop diameter is less than 0.1 cm, the influence of gravity is negligible. The present method and the spherical cap method obtain similar contact angles. This is consistent to the literature[46], in which Picknett and Bexon reported that a droplet resting on a smooth homogeneous surface takes the shape of a spherical cap provided that its mass is less than about 1 mg. When the droplets are larger than this scale, the contact angle computed by the spherical cap method noticeably decreases and increasingly deviates from the values without gravity. However, the contact angle evaluated by the present method keeps almost the same all the time. The droplet deformations are drawn in Fig. 7. The two macroscopic droplet diameters are 0.3 and 0.5 cm on the three kinds of solid surfaces with the chemical potentials 0.2, 0.3 and 0.4, respectively. The surface tensor is constant for a water-vapor system at a given temperature; therefore the deformation of a bigger droplet, which suffers a larger gravity force, is more than that of a smaller one. It can be clearly seen that the droplets with the initial diameter 0.5 cm are flattened much more than those with the initial diameter 0.3 cm. This is consistent to the results of Xie *et al.*[47].

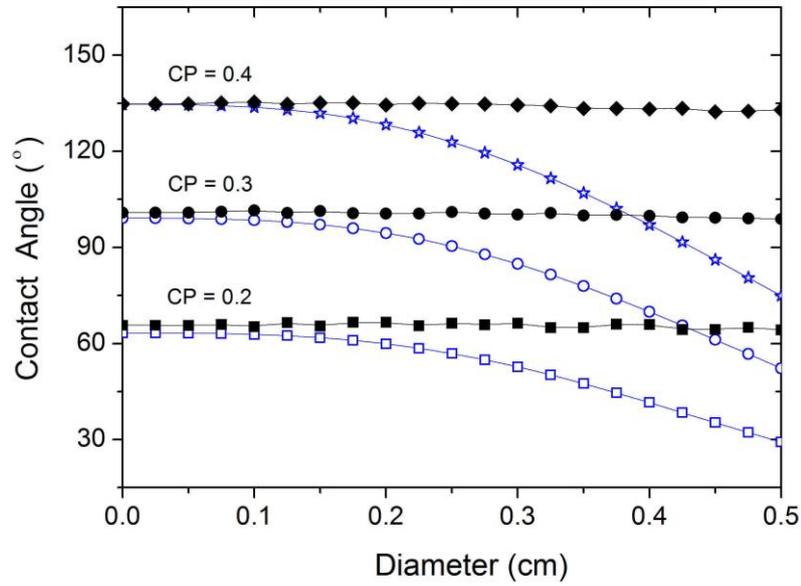

*Fig. 6. (color online) The contact angles of sessile droplets with different diameters on three kind of solid surfaces. The drop diameters change from 0.01 cm to 0.5 cm and the chemical potentials of the solid surfaces are set to 0.2, 0.3 and 0.4, respectively. The black solid symbols are the results of the present method and the blue hollow ones are those from the spherical cap method.*

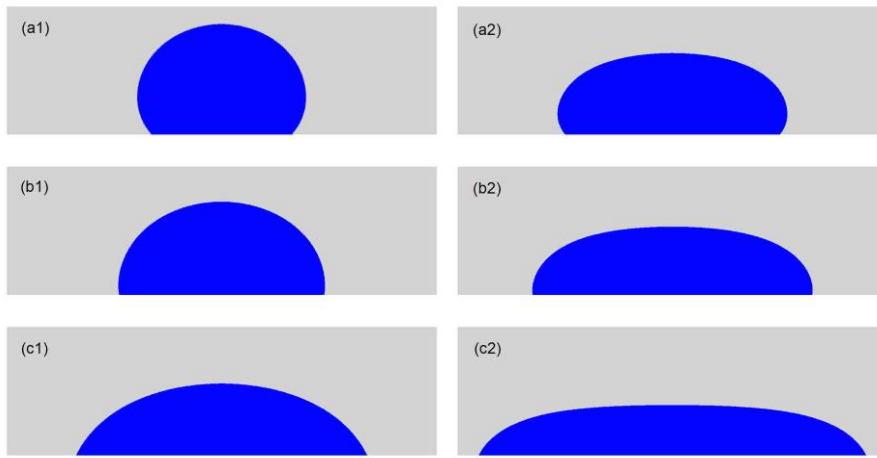

*Fig. 7. (color online) The deformations of sessile droplets with different diameters on three kind of solid surfaces. The initial diameter is 0.3 cm for the left droplets and 0.5 cm for the right droplets. The chemical potentials of the solid surfaces are (a) 0.4, (b) 0.3 and (c) 0.2, respectively.*

### 5.2. Pendent droplets

A pendent droplet is adsorbed on the undersurface of a homogeneous surface. Due to the gravity effect, a pendent droplet will be stretched, its base shrinks and its height will increase. Thus the contact angle calculated by the spherical cap method will increase. We simulate a series of droplets located on three solid undersurfaces with the chemical potentials 0.2, 0.3 and 0.4, respectively. The macroscopic diameters grow gradually and the diameter 0 equals to the situation of zero gravity. Fig. 8 supports that the contact angles evaluated by the present method again remain the same value for droplets with different diameters. When the initial drop diameter is less than 0.1 cm, the influence of gravity is slight and the results from the present method are almost equal to those from the spherical cap method. An apparent distinction between a pendent droplet and a sessile droplet is that a pendent one will drop when its diameter becomes big enough, because the gravity force is increasing along with the growth of the drop size and, finally, it may be larger than the adhesion force. Therefore, as shown in Fig. 8, the drop size cannot increase continuously. The dropping happens when the diameters are larger than 0.40, 0.35 and 0.30 cm for the undersurfaces with the chemical potentials 0.2, 0.3 and 0.4, respectively. The stretching deformations are illustrated in Fig. 9. We draw the largest deformation before dropping, comparing with the smaller droplets. The droplet has a smaller base on a hydrophobic surface than on a hydrophilic one. Therefore, the droplet on the surface with the chemical potential 0.4 is stretched much more and looks more unstable than those on more hydrophilic surface, although it perhaps has a smaller size.

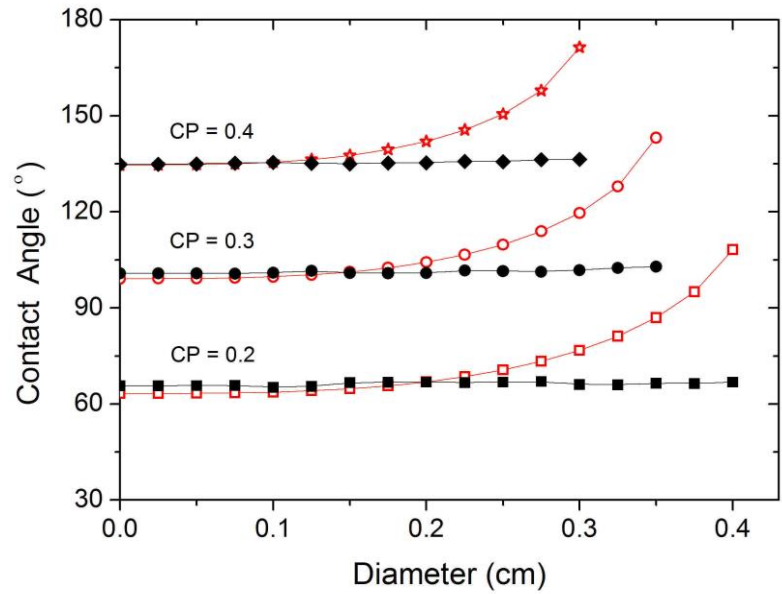

*Fig. 8. (color online) The contact angles of pendent droplets with different diameters on three kind of solid undersurfaces. The drop diameters change from 0.01 cm to 0.5 cm and the chemical potentials of the solid undersurfaces are 0.2, 0.3 and 0.4, respectively. The black solid symbols are the results of the present method and the red hollow ones are those from the spherical cap method.*

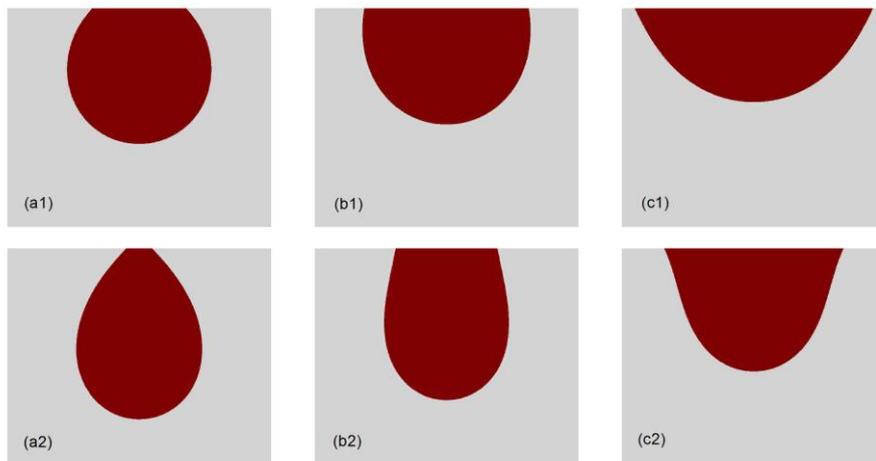

*Fig. 9. (color online) The deformations of pendent droplets with different diameters on three kind of solid undersurfaces. The chemical potentials of the solid undersurfaces are (a) 0.4, (b) 0.3 and (c) 0.2. The initial diameters are (a1) 0.2, (a2) 0.3, (b1) 0.25, (b2) 0.35, (c1) 0.3 and (c2) 0.4 cm.*

## 6. Contact angle hysteresis

In practice, even the cleanest surfaces are not perfectly homogeneous and show chemical or geometrical heterogeneities and these unavoidably lead to contact angle hysteresis[2]. In this section, a droplet on an inclined plate with alternant hydrophobic and hydrophilic patterns is simulated to exhibit contact angle hysteresis on a chemically heterogeneous surface. As shown in Fig. 10, the green segments represent hydrophilic surfaces, while the red segments represent hydrophobic ones. With the effects of gravity and the slope angle $\varphi$, the contact angles of the two sides of the droplet are no longer symmetric and are called advancing and receding contact angles ($\theta_A$ and $\theta_R$), respectively. In this situation, the spherical cap method cannot do anything about the contact angle. The initial drop radius is still $r_0 = 50$ lattice units and its macroscopic diameter is 0.3 cm. The slope angle of the plate increases step by step until the droplet destabilization, at which the advancing or receding contact angles leaves their initial equilibrium locations. Generally, the droplet destabilization is companied with a decrease of the advancing contact angle or an increase of the receding contact angle.

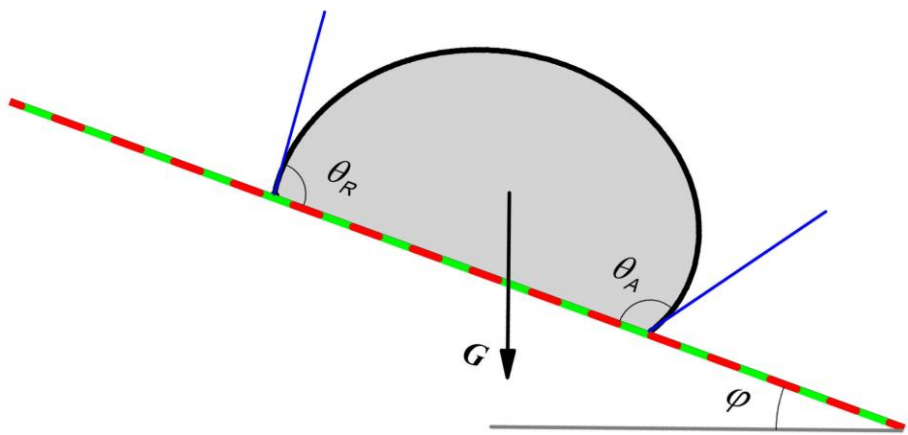

*Fig. 10. (color online) A schematic diagram to illustrate a droplet located on a chemically heterogeneous surface with a slope angle $\varphi$. The segments in green and red represent hydrophilic and hydrophobic surfaces, respectively. With gravity **G**, the droplet displays an advancing contact angle $\theta_A$ and a receding contact angle $\theta_R$.*

The numerical simulations are performed on three kinds of surfaces, which consist of hydrophilic segments with chemical potentials 0.15, 0.20, 0.25 and hydrophobic segments with chemical potentials 0.30, 0.35, 0.40, respectively. Fig. 11 shows the effects of the slope angle of the plate on the contact angles. With the growth of the slope angle, the advancing contact angles gradually increase, while the receding contact angles gradually reduce. These lead to an increasing contact angle hysteresis. Due to the same differences of the hydrophilic and hydrophobic chemical potentials, the change trends of the contact angles on the three surfaces are highly consistent. The more hydrophilic the surface is, the earlier the destabilization happens. Fig. 12 draws the deformation of the droplet on the plate with the slope angles 3º, 6º, 9º and 12º, respectively. The hydrophilic and hydrophobic chemical potentials are 0.25 and 0.40. When the plate is inclined, the advancing contact angle is pushed and enlarged by gravity; meanwhile the receding contact angle is dragged and squeezed by gravity. Therefore, the droplet is askew and the contact angles at the two sides of the droplet are divided into a bigger advancing contact angle and a smaller receding contact angle. The results clearly show that the present contact angle measurement method is valid to capture the dynamic contact angles.

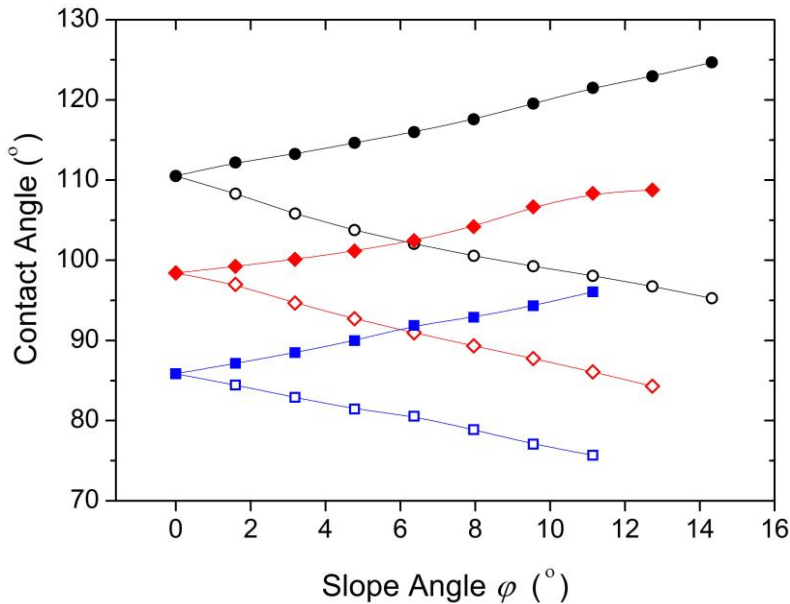

*Fig. 11. (color online) The trends of the contact angles along with the increase of the*

*slope angle of the plate. The symbols represent: ■: $\theta_A$, cp=0.15/0.30; □: $\theta_R$, cp=0.15/0.30; ◆ : $\theta_A$ , cp=0.20/0.35; ◇ : $\theta_R$ , cp=0.20/0.35; ● : $\theta_A$ , cp=0.25/0.40; ○ : $\theta_R$, cp=0.25/0.40.*

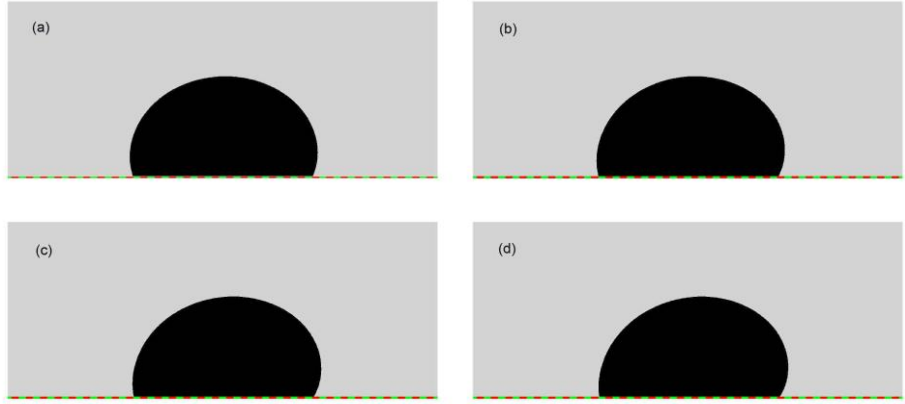

*Fig. 12. (color online) Deformations of droplet on an inclined plate with a chemically heterogeneous surface. The hydrophilic and hydrophobic chemical potentials are 0.25 and 0.40, respectively. The slope angles are (a) $3^o$, (b) $6^o$, (c) $9^o$, (d) $12^o$.*

When the slope angle $\varphi$ is big enough, the destabilized drop will continually slip on the chemically patterned surfaces. The initial drop radius is $r_0 = 60$ lattice units and the macroscopic diameter takes 0.36 cm. A long computational domain with the width 3000 and height 300 lattice units is applied in order that the drop can move long enough for several periods. The contact angles of the hydrophilic and hydrophobic segments are $60^o$ and $120^o$, and the slope angle of the plate is $20^o$. An apparent stick-slip movement is observed and leads to a dynamic contact angle hysteresis. Fig. 13 draws three periods of the drop stick-slip movements. The contact angle hysteresis changes periodically in a range of $5^o$ to $65^o$ and shows two peaks and two troughs in a period. These fluctuations are mainly because the stick-slip movements of the advancing and receding contact angles have a time-phase difference. In Fig. 13 (b) and (c), the lines in light grey are the results of the drop with $r_0 = 100$.

Although there is also a time-phase difference, they are in good agreement with the data from $r_0 = 60$ in a period.

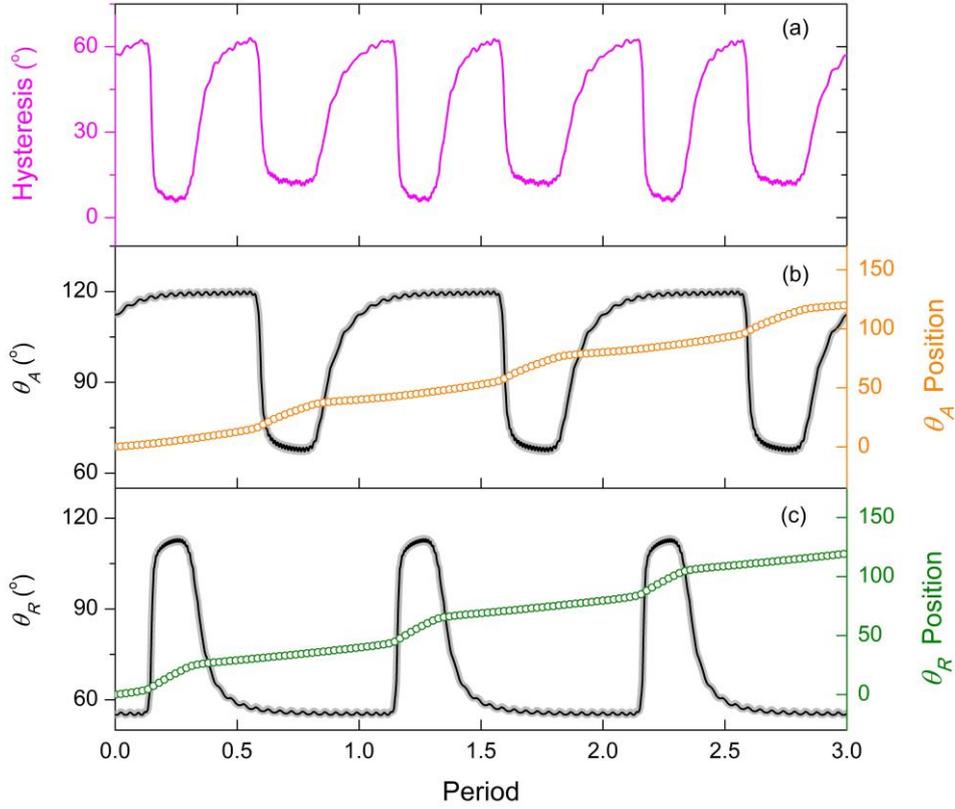

*Fig. 13. (color online) Stick-slip movement of a drop on the chemical patterned surface. The lines in light grey are the results of the drop $r_0 = 100$.*

## 7. Conclusions

Many researches require the spreading of a liquid on a solid surface, which may be clean flat or present some degree of roughness contaminated by compounds with different chemical-physical qualities[12]. Contact angle, which indicates the interactions between solid surface and nonideal fluid, becomes particularly important in these researches. In this paper, we design a geometry-based mesoscopic scheme to measure the contact angle basing on the regular and discrete lattice of the lattice Boltzmann method. The contact angle measurements without the gravity effect are in good agreement with the benchmarks from the spherical cap method. Sessile and pendent droplets on smooth solid surfaces in gravitational environments are simulated and the computational results support that the contact angle is independent of gravity. Furthermore, contact angle hysteresis is simulated on three chemically heterogeneous surfaces. The same differences of the hydrophilic and hydrophobic chemical potentials lead to the similar contact angle hysteresis. These tests fully demonstrate that the proposed contact angle measurement is simple, accurate, robust, and it is suitable for the evaluations of both static and dynamic contact angles. This efficient scheme is expected to promote the on-the-spot measurement of contact angle in dynamic multiphase flow field.

**Acknowledgments**

This work was supported by the National Natural Science Foundation of China (Grant Nos. 11362003, 11462003, 11674345, 11162002), the Key Project of Guangxi Natural Science Foundation (Grant No. 2017GXNSFDA198038), Guangxi Science and Technology Foundation of College and University (Grant No. KY2015ZD017), the Youth Innovation Promotion Association CAS, Guangxi "Bagui Scholar" Teams for Innovation and Research Project, Guangxi Collaborative Innovation Center of Multi-source Information Integration and Intelligent Processing.